\documentclass[prb,twocolumn,showpacs,floatfix,amsmath,amssymb,superscriptaddress]{revtex4}
\usepackage{latexsym}
\usepackage{graphicx}
\usepackage{times}
\usepackage{amsmath}
\usepackage{dcolumn}
\usepackage{latexsym,amsmath,amssymb,bm,euscript}
\newcommand{\ket}[1]{\vert #1 \rangle}

\newcommand{\sgn}{\textrm{sgn}}

\begin{document}
\title{Exotic non-Abelian anyons from conventional fractional quantum Hall states}

\author{David J.~Clarke}
\affiliation{Department of Physics and Astronomy,
University of California, Irvine, CA 92697, USA}

\author{Jason Alicea}
\affiliation{Department of Physics and Astronomy,
University of California, Irvine, CA 92697, USA}

\author{Kirill Shtengel}
\affiliation{Department of Physics and Astronomy,
University of California, Riverside, CA 92521, USA}
\affiliation{Microsoft Research, Station Q, Elings Hall, University of California, Santa Barbara, CA 93106, USA}

\begin{abstract}
Non-Abelian anyons---particles whose exchange noncommutatively transforms a system's quantum state---are widely sought for the exotic fundamental physics they harbor as well as for quantum computing applications.  There now exist numerous blueprints for stabilizing the simplest type of non-Abelian anyon, defects binding Majorana modes, by judiciously interfacing widely available materials.  Following this line of attack, we introduce a device fabricated from conventional fractional quantum Hall states and $s$-wave superconductors that supports exotic non-Abelian anyons that bind `parafermions', which can be viewed as fractionalized Majorana fermions.  We show that these modes can be experimentally identified (and distinguished from Majoranas) using Josephson measurements.  We also provide a practical recipe for braiding parafermions and show that they give rise to non-Abelian statistics.  Interestingly, braiding in our setup produces a richer set of topologically protected qubit operations when compared to the Majorana case.  As a byproduct, we establish a new, experimentally realistic Majorana platform in weakly spin-orbit-coupled materials such as GaAs.
\end{abstract}

\maketitle

\section{Introduction}
\label{Sec:Intro}

The search for non-Abelian anyons in condensed matter has been a focus of both theoretical and experimental efforts in the past decade, driven largely by their potential utility for topological quantum computation~\cite{Nayak2008}. Historically, the first physical system thought to host such exotic quasiparticles was a fractional quantum Hall state at 5/2 filling~\cite{Moore1991}.  A closely related platform for non-Abelian excitations is a chiral $p+ip$ superconductor~\cite{Volovik1999,Read2000}.  While the existence of intrinsic $p+ip$ superconductors in nature remains an open question, Fu and Kane~\cite{Fu2008} provided an important insight by showing that heterostructures formed by a conventional $s$-wave superconductor and a topological insulator effectively mimic the underlying physics and support non-Abelian quasiparticles. Subsequent work showed that the topological insulator can be replaced by a spin-orbit-coupled semiconductor with Zeeman splitting~\cite{Sau2010a,Alicea2010a}. In all of these systems, the non-Abelian anyons arise from Majorana zero-modes bound to vortices~\cite{Volovik1999,Read2000}.  Majorana zero-modes are also predicted to appear at the ends of 1D topological superconducting wires~\cite{Kitaev2001,Lutchyn2010a,Oreg2010}, with recent experimental data supporting such a possibility~\cite{Mourik2012}.
Remarkably, in networks of 1D wires these modes give rise to non-Abelian statistics~\cite{Alicea2011a,Clarke2011a,Halperin2011b}.

The aforementioned candidate systems feature non-Abelian anyons of Ising type, which are \emph{not} computationally universal~\cite{Freedman2002b}. That is, not all unitary quantum gates can be approximated by braiding operations alone. While a number of proposals to circumvent this problem have been put forward~\cite{Bravyi2000-unpublished,Freedman2006,Bonderson2010b}, they are either unrealistic or require non-topological operations, hence weakening the main advantage of using anyons---topological protection from decoherence.  A natural question therefore arises as to whether one can find other types of non-Abelian anyons in realistic condensed matter systems. While there is some hope that the quantum Hall plateau at $\nu = 12/5$ (and perhaps other fillings) may realize non-Abelian anyons with computationally universal braid statistics~\cite{Read1999,Nayak2008}, experimental evidence supporting this idea has yet to appear.

Following the spirit of recent developments in the pursuit of Majorana zero-modes, here we ask whether one can alternatively ``engineer'' a composite system supporting more exotic non-Abelian anyons.  Our search specifically targets quasi-1D platforms for such anyons, similar to Majorana wires.  At first glance, one might be skeptical about the prospects of success here given that Fidkowski and Kitaev~\cite{Fidkowski2011a} showed that the only non-trivial zero-modes supported by 1D electron systems---even with strong interactions---are Majorana fermions.  We will, however, form `wires' out of edge states of topologically nontrivial materials.  The `vacuum' surrounding these `wires' is therefore non-trivial, allowing us to circumvent the aforementioned restrictions.  In particular, we introduce a device employing conventional quantum Hall states (such as at 1/3 filling) and ordinary $s$-wave superconductors that, remarkably, supports `parafermionic' zero-modes that in a sense represent fractionalized Majorana fermions.  These modes generate richer non-Abelian braiding statistics that may render them better candidates for quantum computation, and can be probed via Josephson and tunneling measurements.  Our results open numerous experimentally relevant directions in the search for exotic non-Abelian anyons in condensed matter.

\section{Parafermions from a clock model}
\label{sec-zN}

It is instructive to review a toy model discussed recently by Fendley~\cite{Fendley2012} that supports the parafermion zero-modes that we will later realize in a physical electronic system.  As a primer, let us recall the well-known connection between the transverse field Ising model and a spinless $p$-wave superconductor: the former system maps to the latter under a non-local Jordan-Wigner transformation that trades the bosonic spin variables for fermions.  Although the Ising model exhibits only conventional paramagnetic and ferromagnetic phases, the corresponding superconducting system is far more exotic.  Indeed, while the paramagnetic phase of the former maps to a trivial superconducting state in the latter, the ferromagnetic phase corresponds to a topological state featuring Majorana zero-modes that generate non-Abelian statistics~\cite{Kitaev2001,Alicea2011a,Clarke2011a,Halperin2011b}.

Fendley's crucial insight (guided by the identification of parafermionic fields in the 2D clock model by Fradkin and Kadanoff~\cite{Fradkin1980}) is that one can access still more exotic zero-modes by implementing an analogous non-local transformation on the generalized $N$-state quantum clock model~\cite{Fendley2012},
\begin{equation}
  H = -J\sum_{j = 1}^{L-1}(\sigma^\dagger_j \sigma_{j+1} + H.c.)  - h\sum_{j = 1}^L(\tau_j^\dagger + \tau_j).
  \label{Clock}
\end{equation}
Here $J\geq 0$ couples neighboring `spins' ferromagnetically, $h\geq 0$ is the transverse field, $j$ labels sites of an $L$-site chain, and $\sigma_j, \tau_j$ are operators defined on an $N$-state Hilbert space that satisfy $\sigma_j^N = 1$, $\sigma_j^\dagger = \sigma_j^{N-1}$ and similarly for $\tau_j$.  The only non-trivial commutation relation among these operators reads $\sigma_{j}\tau_j = \tau_j \sigma_{j}e^{2\pi i/N}$.   When $N = 2$ Eq.\ (\ref{Clock}) reduces to the familiar transverse field Ising model, though the phases realized in this special case appear also for general $N$.  For example, with $J = 0$, $h > 0$ there exists a unique paramagnetic ground state with $\tau_j = +1$, while in the $J > 0$, $h = 0$ regime an $N$-fold degenerate ferromagnetic ground state with $\sigma_j = e^{2\pi i q/N}$ emerges ($q = 1,\ldots,N$).

Consider now the non-local transformation~\cite{Fendley2012}
\begin{equation}
  \alpha_{2j-1} = \sigma_j \prod_{i<j}\tau_i,\qquad \alpha_{2j} = -e^{i\pi/N}\tau_j\sigma_j \prod_{i < j} \tau_i.
  \label{alpha}
\end{equation}
The properties of $\sigma_j,\tau_j$ dictate that these new operators satisfy $\alpha_{j}^N = 1$, $\alpha_{j}^\dagger = \alpha_{j}^{N-1}$, and
\begin{equation}
  \alpha_{j}\alpha_{j'} =  \alpha_{j'}\alpha_{j}e^{i\frac{2\pi}{N}\sgn(j'-j)}.
  \label{CommutationRelations}
\end{equation}
Appendix \ref{PFproperties} provides some additional useful relations.  For $N = 2$ the operators $\alpha_j$ are self-Hermitian, anticommute, and square to the identity---hence they are Majorana fermions.  At larger $N$, however, they define \emph{parafermions}~\cite{Fradkin1980,Zamolodchikov1985,Fendley2012}.

\begin{figure}
\includegraphics[width = 7cm]{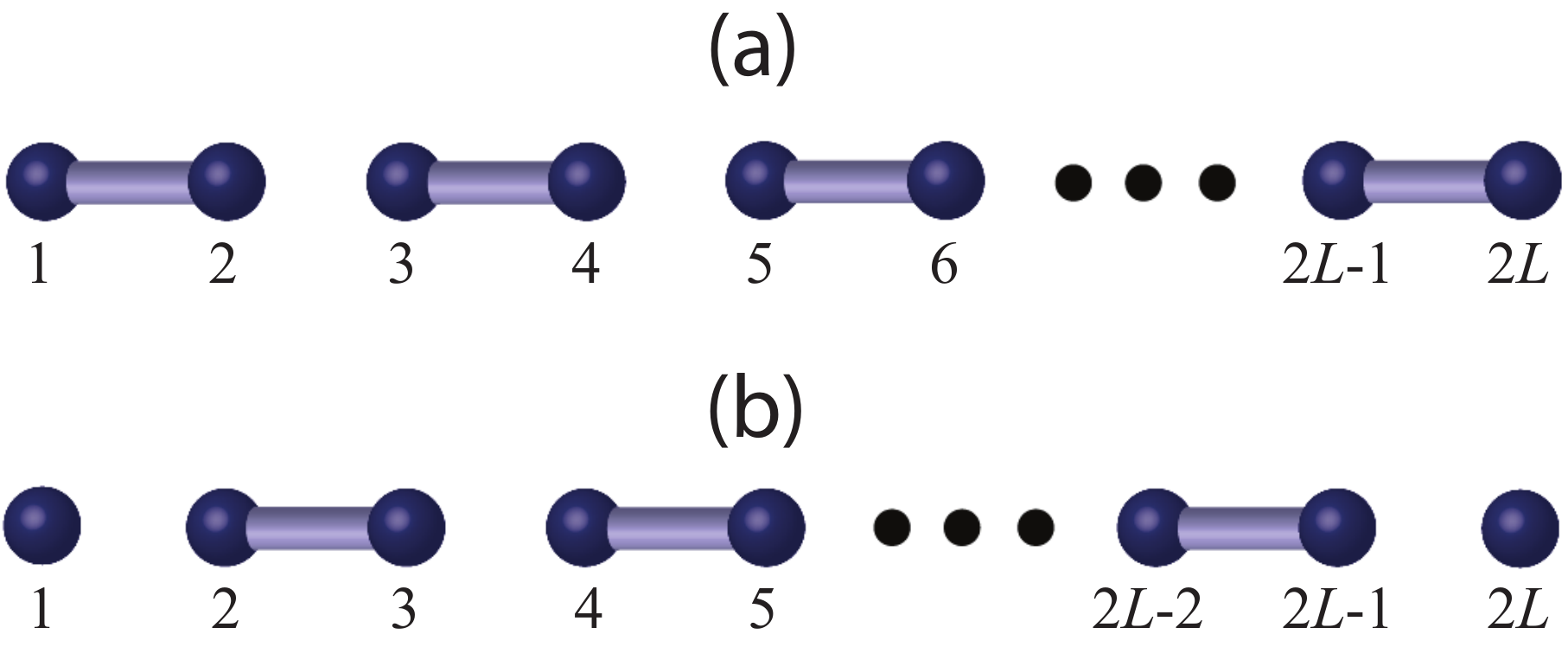}
\caption{Schematic illustration of the parafermion chain Hamiltonian in Eq.\ (\ref{Halpha}) when (a) $J = 0$ and (b) $h = 0$.  In the latter case the ends of the chain support `unpaired' parafermion zero-modes that give rise to an $N$-fold ground-state degeneracy.}
\label{dimers}
\end{figure}

In these variables the Hamiltonian becomes
\begin{eqnarray}
  H &=& J\sum_{j = 1}^{N-1}(e^{-i\frac{\pi}{N}}\alpha_{2j}^\dagger \alpha_{2j+1} + H.c.)
  \nonumber \\
  &+& h\sum_{j = 1}^N(e^{i\frac{\pi}{N}}\alpha_{2j-1}^\dagger \alpha_{2j} + H.c.).
  \label{Halpha}
\end{eqnarray}
The phases of the clock model appear in this representation as follows.  In the paramagnetic limit with $J = 0$ the operators `pair up' as sketched in Fig.\ \ref{dimers}(a)---$\alpha_1$ couples to $\alpha_2$, $\alpha_3$ couples to $\alpha_4$, and so on down the chain.  As shown in Appendix \ref{PFproperties} one can simultaneously diagonalize this collection of `dimers', which take on eigenvalues $\alpha_{2j-1}^\dagger \alpha_{2j} = -e^{i\frac{2\pi}{N}(n_j-1/2)}$ for integer $n_j$.  Here there exists a unique ground state with $n_j = 0$ that is fully gapped since exciting any of these dimers costs finite energy.  More interestingly, the ferromagnetic case $h = 0$ produces the shifted dimerization shown in Fig.\ \ref{dimers}(b).  A bulk gap arises here for the same reason, though the ends of the chain now support `unpaired' zero-modes $\alpha_1$ and $\alpha_{2L}$ that encode the $N$-fold degeneracy of the clock model's ferromagnetic phase ($\alpha_{2L}^\dagger \alpha_{1}$ admits $N$ distinct eigenvalues that do not affect the energy).  At $N = 2$ the zero-mode operators $\alpha_{1,2L}$ form the unpaired Majoranas identified by Kitaev~\cite{Kitaev2001}; for $N>2$ they correspond to parafermion zero-modes\cite{Fendley2012} that are our main interest here.

\section{Practical realization in quantum wells}
\label{sec-phys}

The Majorana zero-modes supported by Eq.\ (\ref{Halpha}) when $N = 2$ are relatively `easy' to engineer, because in this case the operators $\alpha_j$ satisfy familiar fermionic anticommutation relations [see Eq.\ (\ref{CommutationRelations})]. This property allows one to rewrite the $N = 2$ Hamiltonian in terms of ordinary fermion operators $c_j = (\alpha_{2j-1}+i\alpha_{2j})/2$, yielding a model for a spinless $p$-wave superconductor~\cite{Kitaev2001} that can be realized in a variety of experimental architectures~\cite{BeenakkerReview,AliceaReview}.  Because of the non-standard commutation relations obeyed by $\alpha_j$ with $N>2$, however, devising experimental realizations of parafermionic zero-modes is substantially more difficult.  Our approach is inspired by the observation that commutation relations akin to those in Eq.\ (\ref{CommutationRelations}) do occur among physical operators in a familiar system---a fractional quantum Hall edge. In particular, for Laughlin states at filling factor $\nu=1/m$ ($m$ is an odd integer), the quasiparticle operators $e^{i\phi(x)}$ that create right-moving charge $e/m$ excitations at position $x$ along the edge obey~\cite{WenBook}
\begin{equation}
  e^{i\phi(x)}e^{i\phi(x')}=e^{i\phi(x')}e^{i\phi(x)}e^{i\frac{\pi}{m}\sgn(x'-x)}.
  \label{edge_commutation}
\end{equation}
This suggests that such a system may provide a key building block in a device supporting localized parafermion modes.

\begin{figure}
\includegraphics[width = 7.5cm]{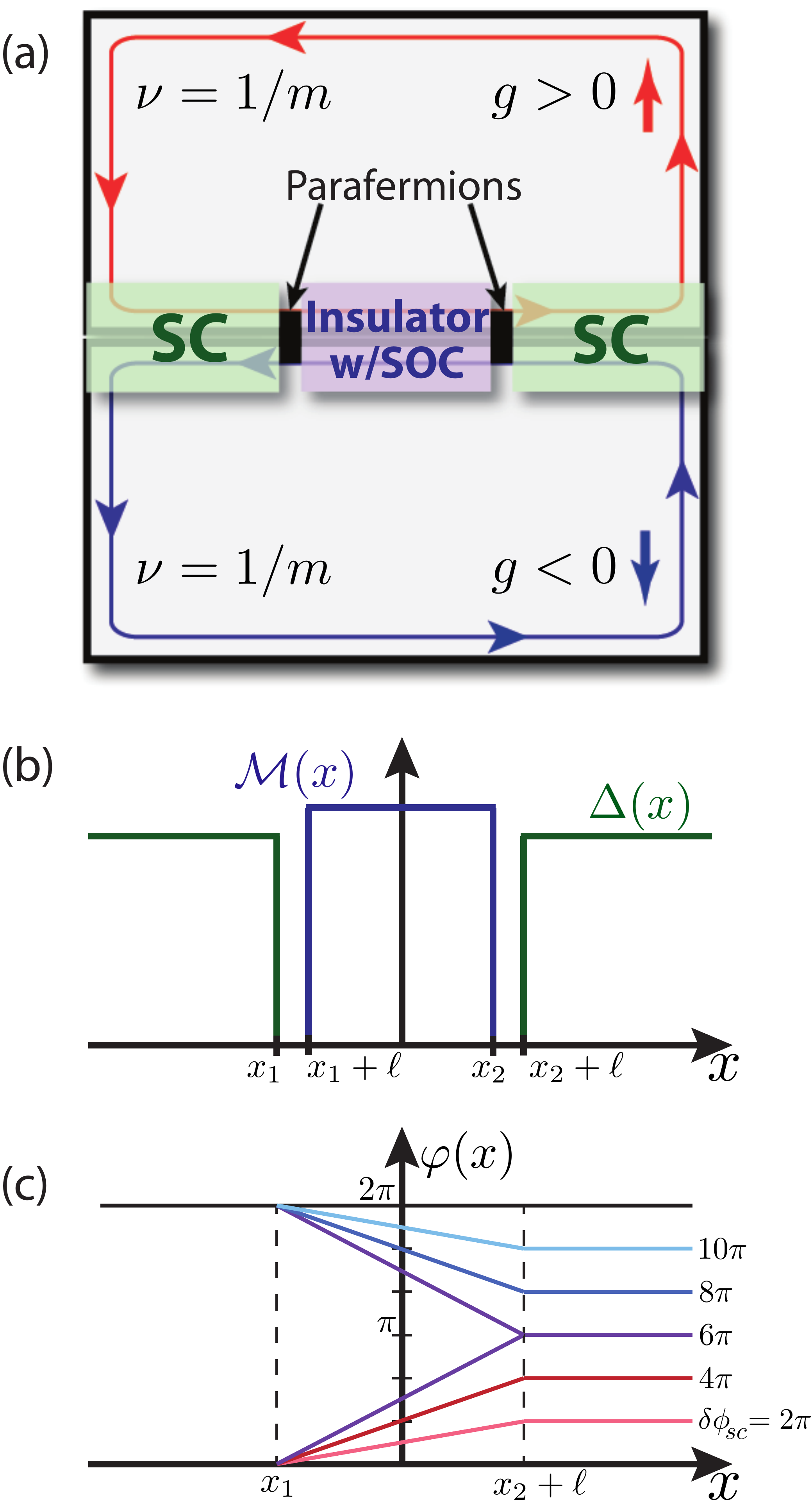}
\caption{(a) Experimental architecture realizing parafermion zero-modes.  (b) Spatial profile for the pairing amplitude $\Delta(x)$ and tunneling strength $\mathcal{M}(x)$ induced by the superconductors and insulator in (a).  (c) Schematic dependence of $\varphi(x)$ on the phase difference $\delta\phi_{sc}$ between the superconductors in (a), in the $m = 3$ case.  As $\delta\phi_{sc}$ winds the larger mismatch between $\varphi(x)$ on the left and right increases the energy until $\delta\phi_{sc} = 6\pi$.  Additional $2\pi$ cycles then `untwist' $\varphi(x)$ until the ground state is again accessed at $\delta\phi_{sc} = 12\pi$.  Remarkably, this implies that the Josepshon current exhibits $12\pi$ periodicity in $\delta\phi_{sc}$.}
\label{FQH_fig}
\end{figure}

A single quantum Hall state is of course insufficient for this purpose, since its edge can not be gapped out (and hence one can not localize modes of any type at the edge).  Thus we consider the geometry of Fig.\ \ref{FQH_fig}(a), where two adjacent quantum wells, each at filling $\nu = 1/m$, give rise to a pair of counterpropagating edge states at their interface.  These modes can acquire a gap via two different mechanisms: first through tunneling of electrons across the junction, and second by assembling electrons from each edge into Cooper pairs.  To facilitate Cooper pair formation, we will assume that the two quantum wells exhibit opposite-sign $g$-factors so that the red and blue edge states in Fig.\ \ref{FQH_fig}(a) carry antiparallel spins.  (In practice one can control the sign of the $g$-factor by various means---see, \emph{e.g.}, Refs.\ \onlinecite{Snelling1991,Malinowski2000}.)  The counterpropagating edge modes are then similar to those of 2D non-interacting topological insulators\cite{Kane2005,Hasan2010,Qi2011} when $m = 1$ and `fractional' topological insulators\cite{Levin2009a} when $m>1$.  This allows a pairing gap to open at the interface via the proximity effect with ordinary $s$-wave superconductors [green regions in Fig.\ \ref{FQH_fig}(a)].  A tunneling gap meanwhile can originate from spin-orbit-induced backscattering between the edge states; the requisite spin-orbit coupling can arise either directly from the quantum wells, or from the insulator [purple region in Fig.\ \ref{FQH_fig}(a)] that electrons traverse when crossing the interface.

Of particular interest are the properties of domain walls between regions gapped by these different means.  One can most simply explore this physics by considering the static domain structure of Fig.\ \ref{FQH_fig}(a), in the limit where electrons tunnel across the interface only via the central spin-orbit-coupled insulator.  (We assume that spin conservation precludes backscattering elsewhere.)  We further postulate that the potential along the edge is tuned so that low energy right- and left-moving $e/m$ quasiparticles carry `small' momenta.  In terms of fields $\phi_{R/L}$ satisfying $[\phi_{R/L}(x),\phi_{R/L}(x')] = \pm i \frac{\pi}{m}\sgn(x-x')$ and $[\phi_L(x),\phi_R(x')] = i\frac{\pi}{m}$, such excitations are created by operators $e^{i\phi_{R/L}}$ that exhibit commutation relations of the form in Eq.\ (\ref{edge_commutation}).  These properties ensure that the electron operators given by $\psi_{R/L} \sim e^{i m \phi_{R/L}}$ obey Fermi statistics.  Below it will prove useful to write $\phi_{R/L} = \varphi \pm \theta$; here $\rho = \partial_x\theta/\pi$ is the electron density operator while
\begin{equation}
  [\varphi(x),\theta(x')] = i\frac{\pi}{m}\Theta(x-x').
  \label{PhiThetaCommutator}
\end{equation}

We model the interface with a Hamiltonian $H = H_0 + H_1$, where\cite{WenBook}
\begin{equation}
  H_0 = \frac{m v}{2\pi}\int dx \left[(\partial_x\varphi)^2+(\partial_x\theta)^2\right]
  \label{H0}
\end{equation}
describes gapless counterpropagating edge modes with speed\footnote{In general the right- and left-moving velocities need not coincide, but our results hold even when these velocities differ.} $v$ and $H_1$ encodes couplings induced by the superconductors and spin-orbit-coupled insulator in Fig.\ \ref{FQH_fig}(a).  (For simplicity we neglect Coulomb interactions between the edge states throughout; also, until specified otherwise we set the superconducting phases to zero.)  In terms of electron operators we have $H_1 = \int dx[\Delta(x)\psi_R\psi_L + \mathcal{M}(x)\psi_R^\dagger \psi_L + H.c.]$, where the profiles for the pairing amplitude $\Delta(x)$ and tunneling strength $\mathcal{M}(x)$ appear in Fig.\ \ref{FQH_fig}(b).  One can alternatively express $H_1$ using $\varphi,\theta$ variables as
\begin{eqnarray}
  H_1 \sim \int dx[-\Delta(x) \cos(2m\varphi) - \mathcal{M}(x)\cos(2m\theta)].
  \label{H1}
\end{eqnarray}
Similar models have been studied previously in the $m = 1$ case, where it is well-established that each domain wall binds a single Majorana fermion\cite{Fu2009a,Sela2011}.  We will now show that, remarkably, for $m>1$ the Hamiltonian supports precisely the localized parafermion zero-modes that we seek.

To this end, consider the limit where the induced pairing and tunneling terms are sufficiently strong that beneath each superconductor $\varphi$ locks to one of the $2m$ minima of the first term in Eq.\ (\ref{H1}), while under the insulator $\theta$ pins to the minima of the second\footnote{Throughout we assume that the \emph{bare} values of the pairing and tunneling are `large' so that a perturbative analysis of their influence does not apply.}.  Using the coordinates specified in Fig.\ \ref{FQH_fig}(b) one can then write $\varphi(x<x_1) = \pi \hat{n}^{(1)}_\varphi/m$, $\theta(x_1+ \ell < x < x_2) = \pi \hat{n}_\theta/m$, and $\varphi(x>x_2+\ell) = \pi \hat{n}^{(2)}_\varphi/m$ where $\hat{n}^{(j)}_\varphi$ and $\hat{n}_\theta$ are integer-valued operators.  Importantly, Eq.\ (\ref{PhiThetaCommutator}) yields
\begin{equation}
  [\hat{n}^{(2)}_\varphi,\hat{n}_\theta] = i\frac{m}{\pi},
  \label{nCommutator}
\end{equation}
whereas the other integer-valued operators commute.   At low energies one can focus on the intervals between $x_{j}$ and $x_{j} + \ell$ in Fig.\ \ref{FQH_fig}(b) where $\Delta(x)$ and $\mathcal{M}(x)$ simultaneously vanish, allowing both $\varphi$ and $\theta$ to fluctuate.  These regions are governed by an effective Hamiltonian
\begin{equation}
  H_\text{eff} = \frac{mv}{2\pi}\sum_{i = 1}^2 \int_{x_i}^{x_i+\ell} dx \left[(\partial_x\varphi)^2+(\partial_x\theta)^2\right],
\end{equation}
subject to boundary conditions imposed by the adjacent gapped regions.  As outlined in Appendix \ref{ZeroModeAppendix}, the operators
\begin{eqnarray}
  \alpha_j &=& e^{i\frac{\pi}{m}(\hat{n}^{(j)}_\varphi+\hat{n}_\theta)}\int_{x_j}^{x_j+\ell}dx[e^{-i\frac{\pi}{m}(\hat{n}^{(j)}_\varphi+\hat{n}_\theta)}e^{i(\varphi + \theta)}
  \nonumber \\
  &+& e^{-i\frac{\pi}{m}(\hat{n}^{(j)}_\varphi-\hat{n}_\theta)}e^{i(\varphi-\theta)}  + {H.c.}].
  \label{ZeroModes}
\end{eqnarray}
commute with $H_\text{eff}$ and thus represent zero-modes bound to the domain walls in Fig.~\ref{FQH_fig}.  Note that $\alpha_j$ alters the charge on the domain wall by $e/m$; the first term simply involves $e^{i\phi_R}$, while the other terms similarly add charge $e/m$.  Apart from these modes $H_\text{eff}$ also supports excitations with a finite-size gap of order $1/\ell$.  We henceforth focus on the ground-state sector where such excitations are absent.  In this limit the zero-mode operators simplify considerably---Appendix \ref{ZeroModeAppendix} demonstrates that upon projecting out gapped excitations of $H_\text{eff}$ the integral in Eq.\ (\ref{ZeroModes}) collapses to an unimportant constant.  One then obtains the elegant expressions
\begin{equation}
  \alpha_{j} \rightarrow  e^{i\frac{\pi}{m}(\hat{n}^{(j)}_\varphi+\hat{n}_\theta)}
  \label{ZeroModesProjected}
\end{equation}
that describe the action of the zero-modes within the ground-state manifold.

From Eqs.\ (\ref{nCommutator}) and (\ref{ZeroModesProjected}) it is straightforward to show that
\begin{equation}
  \alpha_j^{2m} = 1,~~~~ \alpha_j \alpha_{j'} = \alpha_{j'}\alpha_j e^{i\frac{\pi}{m}\sgn(j'-j)}.
  \label{alpha_properties}
\end{equation}
Thus the zero-modes bound to our domain walls indeed provide a physical realization of the parafermions supported by the model discussed in Sec.\ \ref{sec-zN} (with $N = 2m$).  In the present context these modes produce a $2m$-fold ground state degeneracy corresponding to the distinct eigenvalues of $\mathcal{O} \equiv \alpha_1^\dagger \alpha_2 = e^{i\frac{\pi}{m}[\hat{n}^{(2)}_\varphi-\hat{n}^{(1)}_\varphi-1/2]}.$\footnote{Note that sending $\varphi \rightarrow \varphi + \pi/m$, which shifts $\hat{n}^{(1,2)}_\varphi \rightarrow \hat{n}^{(1,2)}_\varphi + 1$, corresponds to a global $Z_2$ gauge transformation $\psi_{R/L}\rightarrow -\psi_{R/L}$; thus only the difference $\hat{n}^{(2)}_\varphi-\hat{n}^{(1)}_\varphi$ is physically meaningful when counting degeneracies.}  One can gain intuition for this result by noting that since $\mathcal{O}^\dagger \hat{n}_\theta \mathcal{O} = \hat{n}_\theta + 1$, $\mathcal{O}$ tunnels between adjacent minima of the $\cos(2m\theta)$ term induced by the spin-orbit-coupled insulator.  The $2m$ distinct eigenstates of $\hat{n}^{(2)}_\varphi-\hat{n}^{(1)}_\varphi$ comprising the ground-state manifold can therefore equivalently be viewed as linear combinations of the $2m$ eigenstates of $\hat{n}_\theta$ characterizing the central tunneling-gapped region.

In the limits studied so far the degeneracy produced by $\alpha_1$ and $\alpha_2$ is exact.  These modes will, however, inevitably hybridize to some degree---due to quasiparticle tunneling between the domain walls---in the more realistic situation where $\varphi$ and $\theta$ are not perfectly pinned by the pairing and tunneling terms.  For instance, tunneling of $e/m$ quasiparticles between the parafermions at lowest order produces a Hamiltonian
\begin{equation}
  \Delta H =  A\alpha_1^\dagger \alpha_2 + H.c. = -|A| \cos\left[\frac{\pi}{m}(\hat{n}_\varphi^{(2)}-\hat{n}_\varphi^{(1)})\right].
\end{equation}
Due to the gap in the region between the parafermions, the coefficient $A$ is exponentially small in $\mathcal{M}$ and the spacing between domain walls. Thus the ground states remain degenerate within exponential accuracy---enabling non-local storage of topological qubits in these states.

Just as for Majorana fermions,\cite{Kitaev2001,Fu2008} the parafermions in our system generate spectacular signatures in Josephson measurements.  To illustrate the physics let us return to the setup of Fig.\ \ref{FQH_fig}(a) in the limit where $\varphi,\theta$ are pinned by the superconductors and insulator.  Suppose that after initializing the system into one of the resulting $2m$ ground states, the tunneling strength $\mathcal{M}(x)$ adiabatically decreases to zero so that the parafermions strongly hybridize between the superconductors.  [We assume that the central region of Fig.\ \ref{FQH_fig}(a) remains gapped by finite-size effects even when $\mathcal{M}(x) = 0$.]  We would like to then understand the Josephson current flowing across the junction when the superconducting phase on the left side remains zero while the phase on the right, $\delta\phi_{sc}$, varies.

The Hamiltonian describing this configuration is $H = H_0 + H_1'$, where $H_0$ is again given in Eq.\ (\ref{H0}) while
\begin{equation}
    H_1'  = -\Delta \!\!\! \int\limits_{x<x_1} \!\!\! dx \cos(2m\varphi)
  - \Delta  \!\!\!\!\! \int\limits_{x>x_2 + \ell}  \!\!\!\!\! dx \cos(2m\varphi-\delta\phi_{sc}).
  \label{H1p}
\end{equation}
Suppose first that when $\delta\phi_{sc} = 0$ we begin in a ground state with $\varphi_< \equiv \varphi(x<x_1) = 0$ and $\varphi_> \equiv \varphi(x>x_2 + \ell) = 0$ (other initial states are examined below).  Upon smoothly increasing $\delta\phi_{sc}$ to $2\pi$ our initial state evolves such that $\varphi_> = \delta\phi_{sc}/{2m}$ to minimize the second term in Eq.\ (\ref{H1p}).  Crucially, this cycle \emph{raises} the system's energy because of the mismatch between $\varphi_<$ and $\varphi_>$; the resulting twist of $\varphi(x)$ between the superconductors costs energy due to the $(\partial_x\varphi)^2$ term in $H_0$.  The system returns to its original state only after $\delta\phi_{sc}$ winds by $4\pi m$, after which one obtains a value $\varphi_> = 2\pi$ that is physically equivalent to our initial value of 0.  Figure \ref{FQH_fig}(c) illustrates this physics in the simplest nontrivial case with $m = 3$.  The Josephson current $I_0(\delta\phi_{sc}) \propto {d\langle H\rangle}/{d\delta\phi_{sc}}$ follows from the energy and hence also admits $4\pi m$ periodicity.  (See Appendix \ref{JosephsonAppendix} for a more quantitative treatment.)

Interestingly, the current-phase relation depends sensitively on the initial state.  Consider now the more general situation where prior to fusing the parafermions across the junction we prepare the system into a ground state characterized by $\hat{n}^{(2)}_\varphi-\hat{n}^{(1)}_\varphi = \delta n$, for some integer $\delta n$.  One can readily generalize the above analysis to show that the current becomes $I_{\delta n}(\delta\phi_{sc}) = I_0(\delta\phi_{sc}+2\pi \delta n)$, which indeed differentiates all physically distinct values of $\delta n$.  Thus the $4\pi m$-periodic Josephson effect both provides a definitive signature of the parafermions in our setup and enables readout of the quantum information they store\cite{Fu2008}.

\section{Parafermion braiding}
\label{sec-trans}

The results from Sec.\ \ref{sec-phys} extend straightforwardly to setups exhibiting arbitrarily many domain walls separating pairing- and tunneling-gapped regions.  In particular, $2\mathcal{N}$ domain walls (numbered sequentially from left to right) localize parafermion zero-modes $\alpha_{1,\ldots,2\mathcal{N}}$ that obey Eq.\ (\ref{alpha_properties}) and produce $(2m)^{\mathcal{N}}$ degenerate ground states.  Next we show that these modes generate non-Abelian statistics.  To do so we will introduce a practical recipe for transporting domain walls and a geometry that permits their meaningful exchange.  As we will see, compared to the Majorana case braiding enables one to perform a richer set of topologically protected operations on the ground state manifold.

\begin{figure}
\includegraphics[width = 8cm]{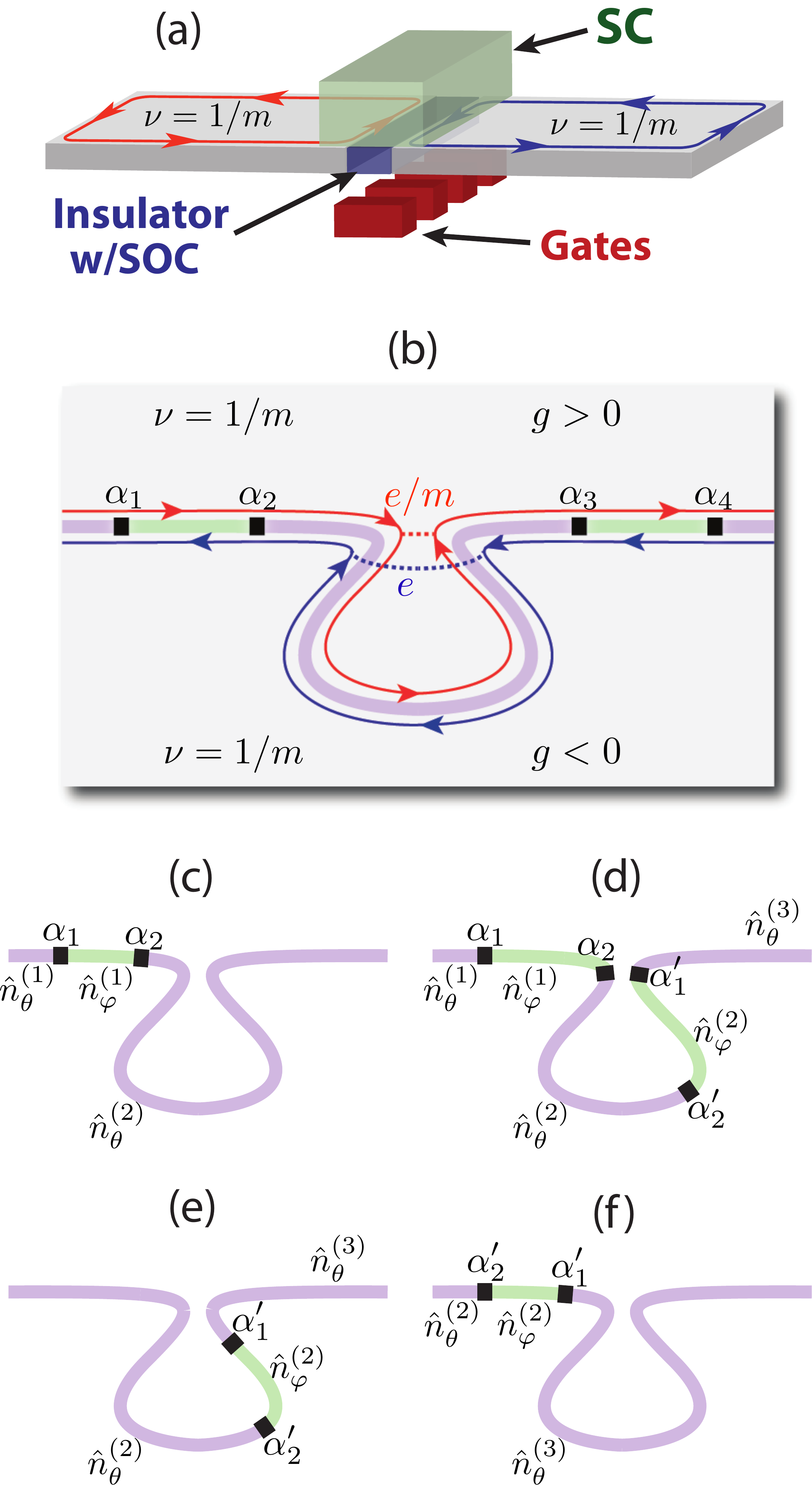}
\caption{(a) Setup allowing adiabatic domain wall transport via gating.  (b) `Sack' geometry that permits braiding of domain walls between tunneling-gapped (purple) and pairing-gapped (green) regions.  Clockwise exchange of domain walls binding $\alpha_1$ and $\alpha_2$ proceeds as outlined in (c)-(f).}
\label{FQH_fig2}
\end{figure}

To mobilize the domain walls we turn now to the alternate setup of Fig.~\ref{FQH_fig2}(a).  Here two quantum wells couple to a superconductor \emph{and} spin-orbit-coupled insulator throughout the interface, while gates below control the potential along the edges.  The Hamiltonian describing the interface is then $H = H_0 + H_1 + H_\mu$, with $H_{0,1}$ given by Eqs.\ (\ref{H0},\ref{H1}) but now with uniform $\mathcal{M}$ and $\Delta$.  The induced potential $\mu(x)$ generates the third term, $H_\mu = -\int dx \mu(x)\partial_x \theta/\pi$ (recall that the electron density is $\rho = \partial_x\theta/\pi$).  We assume $\mathcal{M} \gg \Delta$ so that when $\mu(x) = 0$ a gap arises from inter-edge tunneling.  Suppose that starting from this regime we adjust the gates to raise $\mu$ uniformly.  Shifting $\theta(x) \rightarrow \theta(x) + \frac{\mu}{mv}x$ eliminates the resulting potential term $H_\mu$ but, crucially, also sends $\mathcal{M}\cos(2m\theta)\rightarrow \mathcal{M}\cos(2m\theta + 2\mu x/v)$.  The spatial oscillations render the tunneling term ineffective so that the pairing energy $\Delta$ can then dominate.  More physically, gating changes the momentum carried by low-energy quasiparticles, which affects inter-edge tunneling ($\psi_R^\dagger \psi_L$) but not Cooper pairing ($\psi_R \psi_L$).  The gates in Fig.\ \ref{FQH_fig2}(a) thereby allow one to dynamically control which regions are tunneling- versus pairing-gapped, and hence manipulate domain walls in real time.

Let us now deform the interface into the `sack' geometry sketched in Fig.\ \ref{FQH_fig2}(b), where gates are tuned to localize parafermions $\alpha_{1,\ldots,4}$ (purple and green respectively denote tunneling- and pairing-gapped regions).  This setup permits exchange of any pair of domain walls shown.  We will analyze the adiabatic clockwise braid of the domain walls binding $\alpha_1$ and $\alpha_2$ as outlined in Figs.\ \ref{FQH_fig2}(c)-(f); other exchanges can be understood similarly.  In the figures $\hat{n}_\varphi^{(j)}$ and $\hat{n}_\theta^{(j)}$ denote, as before, integer operators that describe the pinned $\varphi$ and $\theta$ fields in a given region.  Equation (\ref{PhiThetaCommutator}) implies that for $j \geq j'$ $[\hat{n}_\varphi^{(j)},\hat{n}_\theta^{(j')}] = i(m/\pi)$ while for $j < j'$ the operators commute.  Since we are concerned here only with the ground-state sector, it suffices to express $\alpha_1 = e^{i\frac{\pi}{m}(\hat{n}^{(1)}_\varphi+\hat{n}^{(1)}_\theta)}$, $\alpha_2 = e^{i\frac{\pi}{m}(\hat{n}^{(1)}_\varphi+\hat{n}^{(2)}_\theta)}$, and similarly for $\alpha_{1,2}'$.

In the first step [(c)$\rightarrow$(d)] the right domain wall moves into the sack.  We model this process with a Hamiltonian
\begin{eqnarray}
  H_{c\rightarrow d} &=& (t_J\alpha_2^\dagger \alpha_1' + H.c.) + (t\alpha_1'^\dagger \alpha_2' + H.c.)
  \nonumber \\
  &=& -|t_J| \cos\left[\frac{\pi}{m}\left(\hat{n}_\varphi^{(2)} + \hat{n}_\theta^{(3)} -\hat{n}_\varphi^{(1)} - \hat{n}_\theta^{(2)}\right)+\beta\right]
  \nonumber \\
  &-& |t|\cos\left[\frac{\pi}{m}\left(\hat{n}_\theta^{(2)} - \hat{n}_\theta^{(3)}\right)\right].
  \label{Hcd}
\end{eqnarray}
The $t$ term above represents charge $e/m$ tunneling between the domain walls binding $\alpha_1'$ and $\alpha_2'$ in Fig.\ \ref{FQH_fig2}(d).  Similarly, $t_J$ reflects $e/m$ tunneling between the inner (red) edge states at the junction in Fig.\ \ref{FQH_fig2}(b), with $\beta$ a system-dependent phase.  Tunneling can also occur between the outer (blue) edge states, but only of \emph{electrons} since the barrier separating the quantum wells does not support fractionalized excitations.  Electron tunneling has no effect on our conclusions, however, and shall henceforth be neglected.  For details on this important point see Appendix \ref{TunnelingAppendix}.  The configuration in Fig.\ \ref{FQH_fig2}(c) is described by $H_{c\rightarrow d}$ with $t_J = 0$ and $t \neq 0$; $\alpha_1'$ and $\alpha_2'$ then hybridize while $\alpha_{1,2}$ comprise our initial parafermionic zero-modes.  Transporting the right domain wall modifies the Hamiltonian, resulting in $t_J \neq 0$ and $t = 0$ when we arrive at Fig.\ \ref{FQH_fig2}(d).  Throughout this process $\alpha_1$ remains a zero-mode.  The second zero-mode initially given by $\alpha_2$, however, evolves nontrivially.  In principle one can track the latter by explicitly solving $H_{c\rightarrow d}$ for general $t_J,t$---a cumbersome task.

Fortunately, one can deduce the final result far more efficiently by observing that $\chi \equiv e^{i\frac{\pi}{2m}}\alpha_2 \alpha_2'^\dagger \alpha_1' = e^{i\frac{\pi}{m}(\hat{n}_\varphi^{(1)}+\hat{n}_\theta^{(3)})}$ commutes with $H_{c\rightarrow d}$ for any choice of parameters and hence is conserved.  To see how we first note that at the beginning of this step $t$ in Eq.\ (\ref{Hcd}) pins $\hat{n}_\theta^{(3)} = \hat{n}_\theta^{(2)}$.  Consequently $\chi$ projects onto the initial form of the zero-mode of interest, $\chi\rightarrow \alpha_2$.  Upon completing this step $t_J$ instead imposes the energy-minimizing condition
\begin{equation}
  \hat{n}_\varphi^{(2)} + \hat{n}_\theta^{(3)} -\hat{n}_\varphi^{(1)} - \hat{n}_\theta^{(2)} = k,
  \label{constraint}
\end{equation}
for some $\beta$-dependent integer $k$.  Because $\chi$ is conserved, projecting onto the ground-state manifold yields the properly evolved zero-mode:\footnote{Let $U$ be the time-evolution operator and $P_{i/f}$ be the initial/final ground-state projectors.  The zero-mode $\alpha_2$  evolves to $U \alpha_2 U^\dagger = U \chi P_i U^\dagger = \chi U P_i U^\dagger = \chi P_f$.} $\chi \rightarrow e^{i\frac{\pi}{m}(1-k)}\alpha_2'^\dagger \alpha_1'^2$.  Notice that this zero-mode does \emph{not} localize to the transported domain wall, except in the Majorana case where $\alpha_1'^2 = 1$.

Next we transport the domain wall binding $\alpha_1$ to arrive at Fig.\ \ref{FQH_fig2}(e).  This process can be similarly modeled by
\begin{equation}
  H_{d\rightarrow e} = (t_J \alpha_2^\dagger \alpha_1' + H.c.) + (\tilde{t}\alpha_1^\dagger \alpha_2 + H.c.),
\end{equation}
where the parameters change from $t_J \neq 0$, $\tilde t = 0$ at the start of this step to $t_J = 0, \tilde{t} \neq 0$ at the end.  In this case the operator $e^{i\frac{\pi}{m}(1-k)}\alpha_2'^\dagger \alpha_1'^2$ identified above remains a zero-mode throughout, whereas now the zero-mode initially given by $\alpha_1$ evolves.  Proceeding as above, we define $\tilde \chi \equiv e^{-i\frac{\pi}{m}(k-1/2)}\alpha_1\alpha_2^\dagger\alpha_1' = e^{i\frac{\pi}{m}(\hat{n}_\varphi^{(2)}+\hat{n}_\theta^{(3)}-\hat{n}_\theta^{(2)}+\hat{n}_\theta^{(1)}-k)}$ which generically commutes with $H_{d\rightarrow e}$ and initially projects [using Eq.\ (\ref{constraint})] to $\alpha_1$.  When the domain wall moves into the sack $\tilde{t}$ pins $\hat{n}_\theta^{(1)} =  \hat{n}_\theta^{(2)}$.  The zero-mode formerly described by $\alpha_1$ thus evolves to $\tilde\chi \rightarrow e^{i\frac{\pi}{m}(\hat{n}_\varphi^{(2)}+\hat{n}_\theta^{(3)}-k)} = e^{-i\frac{\pi}{m}k}\alpha_1'$ at the end of this step.

During the final step of the exchange, where the domain walls are transported to the configuration in Fig.\ \ref{FQH_fig2}(f), both zero-modes evolve trivially.  Let $U_{12}$ be the operator implementing the exchange.  Comparing Figs.\ \ref{FQH_fig}(c) and (f), one finds that the parafermions transform as
\begin{eqnarray}
  U_{12}\alpha_1U_{12}^\dagger &=& e^{-i \frac{\pi}{m} k}\alpha_2
  \nonumber \\
  U_{12}\alpha_2 U_{12}^\dagger&=& e^{i\frac{\pi}{m}(1- k)}\alpha_1^\dagger\alpha_2^2.
  \label{eq-braid}
\end{eqnarray}
Note that $\alpha_1^\dagger\alpha_2$---the analogue of parity in the Majorana case---is preserved here.  We would like to now understand how the braid transforms the $2m$ degenerate ground states.  Assuming $\alpha_{1,2}$ are the only zero-modes we denote these states by $|q\rangle$ where $\alpha_1^\dagger \alpha_2 |q \rangle = -e^{i\frac{\pi}{m}(q-1/2)}|q\rangle$.  Equations (\ref{eq-braid}) imply that (up to an overall phase)
\begin{equation}
  U_{12} |q\rangle = e^{-i\frac{\pi}{2m}(q-m-k)^2}|q\rangle.
  \label{U12}
\end{equation}
When additional domain walls are present the braids are clearly non-Abelian.  As an example, if $U_{23}$ implements a clockwise exchange of the domain walls binding $\alpha_2$ and $\alpha_3$ in Fig.\ \ref{FQH_fig2}(c), then $U_{23}U_{12}\alpha_1U_{12}^\dagger U_{23}^\dagger=e^{-2\pi ik/m}\alpha_3$ while $U_{12}U_{23}\alpha_1 U_{23}^\dagger U_{12}^\dagger=e^{-\pi i k/m}\alpha_2$.
Appendix \ref{braidapp} discusses additional braids. In particular, there we show that exchanging two pairs of parafermions produces a controlled phase gate $CP=(U_{23}U_{12}U_{34}U_{23})^2$ that can entangle the state of the pair $\alpha_{1}, \alpha_2$ with that of the pair $\alpha_{3}, \alpha_4$. Up to an overall phase, this operation yields
\begin{equation}
  CP\ket{q,q'}=e^{-i\frac{2\pi}{m} (q-k-m)(q'-k-m)}\ket{q,q'},
  \label{CP}
\end{equation}
where $q$ and $q'$ label the eigenvalues of $\alpha_1^\dagger\alpha_2$ and $\alpha_3^\dagger\alpha_4$ respectively. Such an entangling gate is unavailable through braiding of Majorana modes, as can be seen by setting $m=1$, and indicates a relative increase in computational power for parafermions.

\section{Discussion}

In this paper we introduced an experimental setup employing conventional quantum Hall edge states to localize exotic non-Abelian anyons.  In the integer quantum Hall case ($m = 1$) our proposal paves the way towards realizing `Majorana wires' in weakly spin-orbit-coupled systems such as GaAs.  The fractional case $(m >1)$ constitutes a much more important advance, as here our device provides a route to engineering networks of \emph{parafermions} that can be moved along 1D channels.  While we focused on a setup with the virtue of conceptual clarity, numerous variations on the architecture introduced here are possible.  For example, spin-orbit coupling can instead arise from the quantum wells rather than the insulators in Figs.\ \ref{FQH_fig} and \ref{FQH_fig2}.  The opposite-sign $g$-factors we invoked may also be unnecessary, \emph{e.g.}, if coupling to the parent $s$-wave superconductors does not preserve spin~\cite{Chung2011} or if one employs superconductors with a triplet component~\cite{Duckheim2011}.  We have shown that parafermions can be identified---and distinguished from Majoranas---via Josephson measurements.  Tunneling experiments provide a less definitive, though perhaps easier, probe of parafermions.  Indeed, if $\alpha_j$ represents a parafermion zero-mode then $\gamma_j \equiv \alpha_j^m$ is a Majorana operator that can couple to electrons from a lead; thus parafermions give rise to the same quantized zero-bias anomaly as Majoranas~\cite{Law2009}. Experimental verification may also be possible using resonant tunneling of $e/m$ charges in a setting where a parafermion mode localizes inside a quantum Hall point contact (with tunneling charge detectable by, \emph{e.g.}, noise measurements~\cite{Picciotto1997}).

In the future it will be interesting to address whether one can create even more exotic non-Abelian anyons (\emph{e.g.}, Fibonacci) using other fractional quantum Hall states beyond the Laughlin series considered here.  Another worthwhile extension of our work would be to explore three-dimensional fractional topological insulators with proximity-induced superconductivity to generalize Fu and Kane's proposal\cite{Fu2008} for stabilizing Majoranas.  And finally, an important question for future applications is whether one can utilize parafermions to more readily achieve universal quantum computation.  To that end, we note that braiding of Majoranas must be supplemented by two additional gates to achieve universality\cite{Nayak2008}. The first is a single-qubit phase gate, which together with braid transformations completes the set of single-qubit unitary operations. The second is a multi-qubit entangling gate that may be obtained, \emph{e.g.}, by measuring the topological charge of four Majoranas. While the first type of gate remains absent in the parafermion case, the $CP$ braid described above \emph{does} generate the second type. This gate, combined with arbitrary single-`qudit' operations, is universal for quantum computation. We leave to future work an investigation of how such single-`qudit' operations might be realistically performed.

\acknowledgments{We gratefully acknowledge N.\ H.\ Lindner for conversations on independent parallel work [with E.\ Berg, G.\ Refael, and A.\ Stern].  We also thank P.~Fendley for openly sharing unpublished results with us; P.\ Bonderson, J.~P.~Eisenstein, L.\ Fidkowski, M.~P.~A.~Fisher, A.\ Kitaev, and Z.\ Wang for helpful discussions; and Microsoft Station Q and Caltech for hospitality.  This work was supported by the National Science Foundation through grants DMR-1055522 (DC and JA) and DMR-0748925 (KS), the Alfred P.\ Sloan Foundation (JA), and the DARPA-QuEST program (DC and KS). }

\appendix

\section{Properties of parafermion operators}
\label{PFproperties}

Here we will enumerate several useful properties of parafermion operators that follow from the definitions provided in Sec.\ \ref{sec-zN}.  Since $\alpha_j^N = I$ and $\alpha_j^\dagger = \alpha_j^{N-1}$, these operators are unitary and exhibit eigenvalues of the form $e^{2\pi i q/N}$ for integral $q$.  The second equation together with the commutation relations in Eq.\ (\ref{CommutationRelations}) imply that
\begin{equation}
  \alpha_{j}^\dagger\alpha_{j'} =  \alpha_{j'}\alpha_{j}^\dagger e^{-\frac{2\pi i}{N}\sgn(j'-j)}.
  \label{CommutationRelations2}
\end{equation}
Moving $\alpha_{j'}$ past $\alpha_j$ therefore produces the opposite phase factor compared to moving $\alpha_{j'}$ past $\alpha_j^\dagger$.  Consequently we obtain the following commutation relations,
\begin{equation}
  \left[\alpha_i^\dagger \alpha_j,\alpha_k\right] = \left[\alpha_i^\dagger \alpha_j,\alpha_k^\dagger\right] = 0~~~~(k>i,j ~ \text{or}~k < i,j),
  \label{CommutationRelations3}
\end{equation}
which further imply that
\begin{equation}
  \left[\alpha_i^\dagger \alpha_j,\alpha_k^\dagger \alpha_l\right] = 0
  \label{CommutationRelations4}
\end{equation}
so long as neither $k$ nor $l$ lie between $i$ and $j$. 

Equations (\ref{CommutationRelations3}) and (\ref{CommutationRelations4}) demonstrate that as claimed in Sec.\ \ref{sec-zN} one can indeed simultaneously diagonalize each of the `dimers' sketched in Figs.\ \ref{dimers}(a) and (b), as well as the combination of zero-mode operators $\alpha_{2L}^\dagger \alpha_1$ in case (b).  To deduce the allowed eigenvalues we note that one can show from the properties above that
\begin{equation}
  \left(\alpha_i^\dagger \alpha_j\right)^N = (-1)^{N+1},
\end{equation}
which constrains the eigenvalues of $\alpha_i^\dagger \alpha_j$ to the form $-e^{i\frac{2\pi}{N}(q-1/2)}$, where $q$ is an integer.  In the quantum clock model context, the eigenvalues of the relevant `dimer' operators can alternatively be found using the relations
\begin{eqnarray}
  \alpha_{2j-1}^\dagger \alpha_{2j} &=& -e^{-i\pi/N}\tau_j
  \nonumber \\
  \alpha_{2j}^\dagger \alpha_{2j+1} &=& -e^{i\pi/N}\sigma_j^\dagger \sigma_{j+1}
  \label{alphaRelations} \\
  \alpha_{2L}^\dagger \alpha_1 &=&  -e^{i\pi/N} \left(\prod_{i = 1}^L \tau_i^\dagger\right)\sigma_L^\dagger \sigma_1
  \nonumber
\end{eqnarray}
that arise from the non-local transformation specified in Eqs.\ (\ref{alpha}).  Equations (\ref{alphaRelations}) yield the same eigenvalue spectrum for the operators on the left-hand side as noted above since $\tau_j$ and $\sigma_j$ both exhibit non-degenerate eigenvalues $e^{2\pi iq/N}$ for $q = 1,\ldots,N$.

Finally, we consider the case where $\alpha_1$ and $\alpha_{2L}$ represent zero-modes and deduce the action of these operators on the ground state manifold.  Let $|q\rangle$ be a ground state satisfying $\alpha_1^\dagger \alpha_{2L}|q\rangle = -e^{i\frac{2\pi}{N}(q-1/2)}|q\rangle$.  Using the parafermion commutation relations one can show that
\begin{equation}
  (\alpha_1^\dagger \alpha_{2L})\alpha_j = e^{-i\frac{2\pi}{N}}\alpha_j(\alpha_1^\dagger\alpha_{2L})
\end{equation}
for either $j = 1$ or $2L$.  This equation implies that $\alpha_{1,2L}^\dagger|q\rangle \propto |q+1\rangle$ and $\alpha_{1,2L}|q\rangle \propto |q-1\rangle$, where the proportionality constants have unit magnitude.  One can always fix the relative phases of the ground states such that
\begin{equation}
  \alpha_1^\dagger|q\rangle = |q+1\rangle,~~~~\alpha_1|q\rangle = |q-1\rangle.
\end{equation}
With this convention $\alpha_{2L}$ then acts as follows:
\begin{eqnarray}
  \alpha_{2L}^\dagger|q\rangle &=& -e^{-i\frac{2\pi}{N}(q-1/2)}|q+1\rangle
  \nonumber \\
  \alpha_{2L}|q\rangle &=& -e^{i\frac{2\pi}{N}(q-1/2)}|q-1\rangle.
\end{eqnarray}

\section{Solution of localized parafermion zero-modes}
\label{ZeroModeAppendix}

This Appendix provides a detailed derivation of the localized parafermion zero-mode operators quoted in the main text [Eq.\ (\ref{ZeroModes})]. Consider again the static domain structure of Fig.\ \ref{FQH_fig}(a).  As in the main text we will assume that the Cooper pairing and inter-edge tunneling terms induced at the interface are sufficiently strong that $\varphi(x)$ is pinned beneath the superconductors while $\theta(x)$ is pinned beneath the spin-orbit-coupled insulator.  In the black regions of width $\ell$ in Fig.\ \ref{FQH_fig}(a), however, both fields can fluctuate since the pairing and tunneling terms simultaneously vanish there.  Our objective is to now understand the low-energy properties of these regions, which will eventually lead us to the parafermionic zero-mode operators of interest.

For clarity we examine each domain wall separately, beginning with the one on the left.  At low energies this domain wall is governed by an effective Hamiltonian
\begin{equation}
  H = \frac{m v}{2\pi}\int_{x_1}^{x_1+\ell} dx \left[(\partial_x\varphi)^2 + (\partial_x\theta)^2\right].
  \label{Hfree}
\end{equation}
Minimizing the energy of the adjacent gapped regions requires that these fields also satisfy boundary conditions $\varphi(x = x_1) = \pi \hat{n}^{(1)}_\varphi/m$ and $\theta(x = x_1+\ell) = \pi \hat{n}_\theta/m$ for integer-valued operators $\hat{n}_{\varphi}$ and $\hat{n}_\theta$ [recall Eq.\ (\ref{H1})].  Note that $\hat{n}^{(1)}_\varphi$ and $\hat{n}_\theta$ commute since $\left[\varphi(x_1),\theta(x_1+\ell)\right] = 0$ according to Eq.\ (\ref{PhiThetaCommutator}).

Equation (\ref{Hfree}) can be diagonalized by expanding the $\varphi$ and $\theta$ fields as
\begin{eqnarray}
  \varphi(x)&=&\frac{\pi\hat{n}^{(1)}_\varphi}{m}
+\sqrt{\frac{2}{m}}\sum_{k=0}^\infty\frac{\sin\lambda_k(x)}{\sqrt{2k+1}}i
\left(a_k-a_k^\dagger\right),
  \nonumber \\
\theta(x)&=&\frac{\pi\hat{n}_\theta}{m}
+\sqrt{\frac{2}{m}}\sum_{k=0}^\infty\frac{\cos\lambda_k(x)}{\sqrt{2k+1}}
\left(a_k+a_k^\dagger\right),
\label{PhiThetaExpansion}
\end{eqnarray}
where $\lambda_k(x) = \frac{(2k+1)\pi (x-x_1)}{2\ell}$ and $a_{k}$ correspond to conventional bosonic operators satisfying $[a_k,a_{k'}^\dagger] = \delta_{k,k'}$.  [This decomposition simultaneously preserves the commutation relations amongst $\varphi(x), \theta(x')$ and encodes the boundary conditions specified above.]  Inserting Eqs.\ (\ref{PhiThetaExpansion}) into the Hamiltonian yields
\begin{eqnarray}
  H &=& \sum_{k=0}^\infty\epsilon_k(a_k^\dagger a_k + 1/2)
  \label{H_interface}
  \\
  \epsilon_k &=& \frac{\pi v}{\ell}(k+1/2).
\end{eqnarray}
Thus we see that the $a_k$ bosons exhibit a finite-size gap inversely proportional to $\ell$.  This does not, however, necessarily imply that the domain wall admits only gapped excitations.  We will now demonstrate that, because the operators $\hat{n}^{(1)}_\varphi$ and $\hat{n}_\theta$ both commute with the Hamiltonian, the domain wall supports zero-modes that can be constructed from local operators present in our edge theory.

To do so it is convenient to work with chiral fields $\phi_{R/L}$ instead of $\varphi,\theta$.  From Eqs.\ (\ref{PhiThetaExpansion}) we have
\begin{equation}
  \phi_{R/L}(x) = \frac{\pi}{m}\left(\hat{n}^{(1)}_\varphi \pm \hat{n}_\theta\right) \pm \sqrt{\frac{2}{q}}\sum_{k = 0}^\infty\frac{[e^{\pm i\lambda_k(x)}a_k + H.c.]}{\sqrt{2k+1}},
  \label{ChiralFieldExpansion}
\end{equation}
which yields the useful relations
\begin{eqnarray}
  \phi_{L}(x_1) &=& \frac{2\pi}{m}\hat{n}^{(1)}_\varphi-\phi_R(x_1)
  \nonumber \\
  \phi_L(x_1 + \ell) &=& -\frac{2\pi}{m}\hat{n}_\theta+\phi_R(x_1+\ell).
  \label{phiRLrelation}
\end{eqnarray}
One can employ Eqs.~(\ref{H_interface}) and (\ref{ChiralFieldExpansion}) to show, with the aid of various commutator identities and some algebra, that
\begin{equation}
  \left[H,e^{i\phi_{R/L}(x)}\right] = \pm i v \partial_x e^{i\phi_{R/L}(x)}.
\end{equation}
Since the right-hand side is a total derivative the above equation integrates to
\begin{equation}
  \left[H,\int_{x_1}^{x_1+\ell} dx e^{i\phi_{R/L}(x)}\right] = \pm i v \left(e^{i\phi_{R/L}(x_1 + \ell)} - e^{i\phi_{R/L}(x_1)}\right)
  \label{Hcommutator}
\end{equation}
Finally, Eqs. (\ref{phiRLrelation}) and (\ref{Hcommutator}) allow one to demonstrate that the operator
\begin{eqnarray}
  \alpha_1 &=& e^{i\frac{\pi}{m}(\hat{n}^{(1)}_\varphi+\hat{n}_\theta)}\int_{x_1}^{x_1+\ell}dx[e^{-i\frac{\pi}{m}(\hat{n}^{(1)}_\varphi+\hat{n}_\theta)}e^{i\phi_R(x)}
  \nonumber \\
  &+& e^{-i\frac{\pi}{m}(\hat{n}^{(1)}_\varphi-\hat{n}_\theta)}e^{i\phi_L(x)}  + H.c.]
  \label{alpha1}
\end{eqnarray}
commutes with the Hamiltonian and therefore represents a zero-mode of the system.

Several points are worth emphasizing here.  First, $\alpha_1$ is constructed purely from local $e/m$ quasiparticle operators $e^{i\phi_{R/L}(x)}$ and thus represents a physical zero-mode bound to the domain wall.  To see this explicitly, observe that $\hat{n}^{(1)}_\varphi$ and $\hat{n}_\theta$ always appear in $\alpha_1$ via $e^{i\frac{2\pi}{m}\hat{n}^{(1)}_\varphi} = e^{i[\phi_R(x_1) + \phi_L(x_1)]}$ and $e^{i\frac{2\pi}{m}\hat{n}_\theta} = e^{i[\phi_R(x_1+\ell) - \phi_L(x_1+\ell)]}$.  Second, the expression for the zero-mode quoted above is not unique; one can always multiply $\alpha_1$ by allowed operators (such as $H$ or $e^{-i\frac{2\pi}{m}\hat{n}_\theta}$) that also commute with the Hamiltonian.  This freedom---which is a generic feature of zero-mode operators---is, however,  inconsequential for our purposes.  We are concerned here only with physics of the ground-state manifold and in operators that cycle the system across the various ground states, and for this purpose the form of $\alpha_1$ above suffices.  Third, note from Eq.\ (\ref{ChiralFieldExpansion}) that the integrand in Eq.\ (\ref{alpha1}) involves only bosonic operators $a_k,a_k^\dagger$ (and not $\hat{n}^{(1)}_\varphi$ or $\hat{n}_\theta$).  This makes the projection of $\alpha_1$ into the ground-state manifold with $a_k^\dagger a_k = 0$ very simple; upon discarding an overall constant one obtains the result
\begin{equation}
  \alpha_1 \rightarrow e^{i\frac{\pi}{m}(\hat{n}^{(1)}_\varphi+\hat{n}_\theta)}
\end{equation}
quoted in the main text.

The right domain wall in Fig.\ \ref{FQH_fig}(a) can be analyzed very similarly.  Here the effective low-energy Hamiltonian is given by
\begin{equation}
  H = \frac{m v}{2\pi}\int_{x_2}^{x_2+\ell} dx [(\partial_x\varphi)^2 + (\partial_x\theta)^2],
  \label{Hfree3}
\end{equation}
where now the fields satisfy boundary conditions $\theta(x_2) = \pi \hat{n}_\theta/m$ and $\varphi(x_2 + \ell) = \pi \hat{n}^{(2)}_\varphi/m$.  [Since the left and right domain walls of Fig.\ \ref{FQH_fig}(a) are bridged by a single spin-orbit-coupled insulator, $\theta(x_1+\ell)$ and $\theta(x_2)$ must be pinned to identical values.  Thus $\hat{n}_\theta$ is the same integer-valued operator that we introduced above.  However, $\varphi(x_1)$ and $\varphi(x_2 + \ell)$ are pinned by different superconductors, which necessitates the introduction of distinct integer-valued operators $\hat{n}^{(1,2)}_\varphi$.]  Importantly, in this geometry $\hat{n}^{(2)}_\varphi$ and $\hat{n}_\theta$ no longer commute since Eq.\ (\ref{PhiThetaCommutator}) yields $[\varphi(x_2+\ell),\theta(x_2)] = (\pi/m)^2[\hat{n}^{(2)}_\varphi, \hat{n}_\theta] = i\pi/m$.

Given our new boundary conditions, the appropriate decomposition for $\varphi$ and $\theta$ reads
\begin{eqnarray}
  \varphi(x) &=& \frac{\pi}{m}\hat{n}^{(2)}_\varphi + \sqrt{\frac{2}{m}}\sum_{k=0}^\infty\frac{\cos\lambda'_k(x)}{\sqrt{2k+1}}(a_k+a_k^\dagger)
  \nonumber \\
  \theta(x) &=& \frac{\pi\hat{n}_\theta}{m} + \sqrt{\frac{2}{m}}\sum_{k=0}^\infty\frac{\sin\lambda'_k(x)}{\sqrt{2k+1}}i(a_k-a_k^\dagger),
  \label{PhiThetaExpansion2}
\end{eqnarray}
with $\lambda'_k(x) = \frac{(2k+1)\pi (x-x_2)}{2\ell}$ and $[a_k,a_{k'}^\dagger] = \delta_{k,k'}$ as before.  With this expansion our low-energy Hamiltonian once again takes the form in Eq.\ (\ref{H_interface}).  One can then follow the steps outlined above (taking care to enforce the non-trivial commutation relations between $\hat{n}^{(2)}_\varphi$ and $\hat{n}_\theta$) to show that the right domain wall binds a zero-mode described by an operator
\begin{eqnarray}
  \alpha_2 &=& e^{i\frac{\pi}{m}(\hat{n}^{(2)}_\varphi+\hat{n}_\theta)}\int_{x_2}^{x_2+\ell}dx[e^{-i\frac{\pi}{m}(\hat{n}^{(2)}_\varphi+\hat{n}_\theta)}e^{i\phi_R(x)}
  \nonumber \\
  &+& e^{-i\frac{\pi}{m}(\hat{n}^{(2)}_\varphi-\hat{n}_\theta)}e^{i\phi_L(x)}  + H.c.].
  \label{alpha2}
\end{eqnarray}
Projecting onto the ground-state manifold yields, up to a constant,
\begin{equation}
  \alpha_2 \rightarrow e^{i\frac{\pi}{m}(\hat{n}^{(2)}_\varphi+\hat{n}_\theta)}.
\end{equation}

\section{Parafermion Josephson effect}
\label{JosephsonAppendix}

In the main text we deduced the qualitative dependence of the energy, and hence the Josephson current, on the phase difference $\delta\phi_{sc}$ between the two superconductors in Fig.\ \ref{FQH_fig}(a).  This Appendix explores the physics uncovered there more quantitatively.  We continue to work in the limit where the tunneling strength $\mathcal{M}(x)$ vanishes whereas the pairing fields in Eq.\ (\ref{H1p}) pin $\theta$ beneath each superconductor.  The normal region between the superconductors can then be described by an effective Hamiltonian
\begin{equation}
  H = \frac{m v}{2\pi}\int_{x_1}^{x_2+\ell} dx [(\partial_x\varphi)^2 + (\partial_x\theta)^2],
  \label{Hfree2}
\end{equation}
subject to boundary conditions on $\varphi(x_1)$ and $\varphi(x_2 + \ell)$ induced by the neighboring superconductors.

Because (in contrast to Appendix \ref{ZeroModeAppendix}) the same field is now pinned at both endpoints, it is essential that one incorporates compactness of $\varphi$ in what follows; failure to do so yields incorrect results for the dependence of the energy on $\delta\phi_{sc}$.  We will for simplicity set $\varphi(x_1) = 0$---that is, we fix the eigenvalue of the operator $\hat{n}^{(1)}_\varphi$ defined earlier to zero without loss of generality.  At the right boundary, however, we take
\begin{equation}
  \varphi(x_2 + \ell) = \text{mod}\left[ \frac{\pi}{m}\left(\hat{n}^{(2)}_\varphi + \frac{\delta\phi_{sc}}{2\pi}\right) + \pi ,2\pi\right]-\pi,
  \label{varphiBC}
\end{equation}
where $\hat{n}^{(2)}_\varphi$ is the same integer-valued operator introduced previously.  The right-hand side of Eq.\ (\ref{varphiBC}) minimizes the pairing term in Eq.\ (\ref{H1p}) and importantly also restricts $\varphi(x_2 + \ell)$ to lie between $-\pi$ and $\pi$ for any $\delta\phi_{sc}$ and $\hat{n}^{(2)}_\varphi$.  Imposing this bound on the range of $\varphi(x_2 + \ell)$ ensures that $\varphi(x)$ need not exhibit any unnecessary twists between $x = x_1$ and $x_2 + \ell$.

To diagonalize the Hamiltonian we decompose $\varphi,\theta$ as follows:
\begin{eqnarray}
  \varphi(x) &=& \varphi(x_2 + \ell)\frac{x-x_1}{x_2+\ell-x_1}
  \nonumber \\
  &+& \frac{1}{\sqrt{m}}\sum_{k = 1}^\infty \frac{\sin\tilde\lambda_k(x)}{\sqrt{k}} i(a_k-a_k^\dagger)
  \label{varphiDecomposition}
  \\
  \theta(x) &=& \theta_0 + \frac{1}{\sqrt{m}} \sum_{k = 1}^\infty \frac{\cos\tilde\lambda_k(x)}{\sqrt{k}}(a_k + a_k^\dagger),
  \label{thetaDecomposition}
\end{eqnarray}
where $\tilde\lambda_k(x) = \frac{k \pi(x-x_1)}{x_2+\ell-x_1}$ and as usual $a_k$ are canonical bosons satisfying $[a_k,a_{k'}^\dagger] = \delta_{k,k'}$.   In Eq.\ (\ref{thetaDecomposition}) $\theta_0$ represents the zero-momentum component of $\theta(x)$ (note that $k = 0$ is excluded from both sums above).  The boundary conditions on $\varphi(x)$ are clearly obeyed in this representation, while the commutation relations between $\varphi,\theta$ in Eq.\ (\ref{PhiThetaCommutator}) are also preserved provided $\theta_0$ and $\hat{n}^{(2)}_\varphi$ are conjugate variables satisfying $[\hat{n}^{(2)}_\varphi,\theta_0] = i$.  Using the decomposition in Eqs.\ (\ref{varphiDecomposition}) and (\ref{thetaDecomposition}) one can express the effective Hamiltonian as
\begin{eqnarray}
  H &=& \sum_{k = 1}^\infty \tilde\epsilon_k(a_k^\dagger a_k + 1/2) + \mathcal{E}(\delta\phi_{sc})
  \\
  \mathcal{E}(\delta\phi_{sc}) &=& \frac{m v}{2\pi}\frac{[\varphi(x_2 + \ell)]^2}{x_2 + \ell-x_1},
  \label{E}
\end{eqnarray}
with $\varphi(x_2 + \ell)$ given by Eq.\ (\ref{varphiBC}).  The first term in $H$ above simply describes gapped excitations with energy $\tilde\epsilon_k = \frac{\pi v}{x_2 + \ell-x_1}k$, which we assume are absent.  More interestingly, the second term captures the dependence of the energy on the superconducting phase difference imposed across the junction.

\begin{figure}
\includegraphics[width = 8cm]{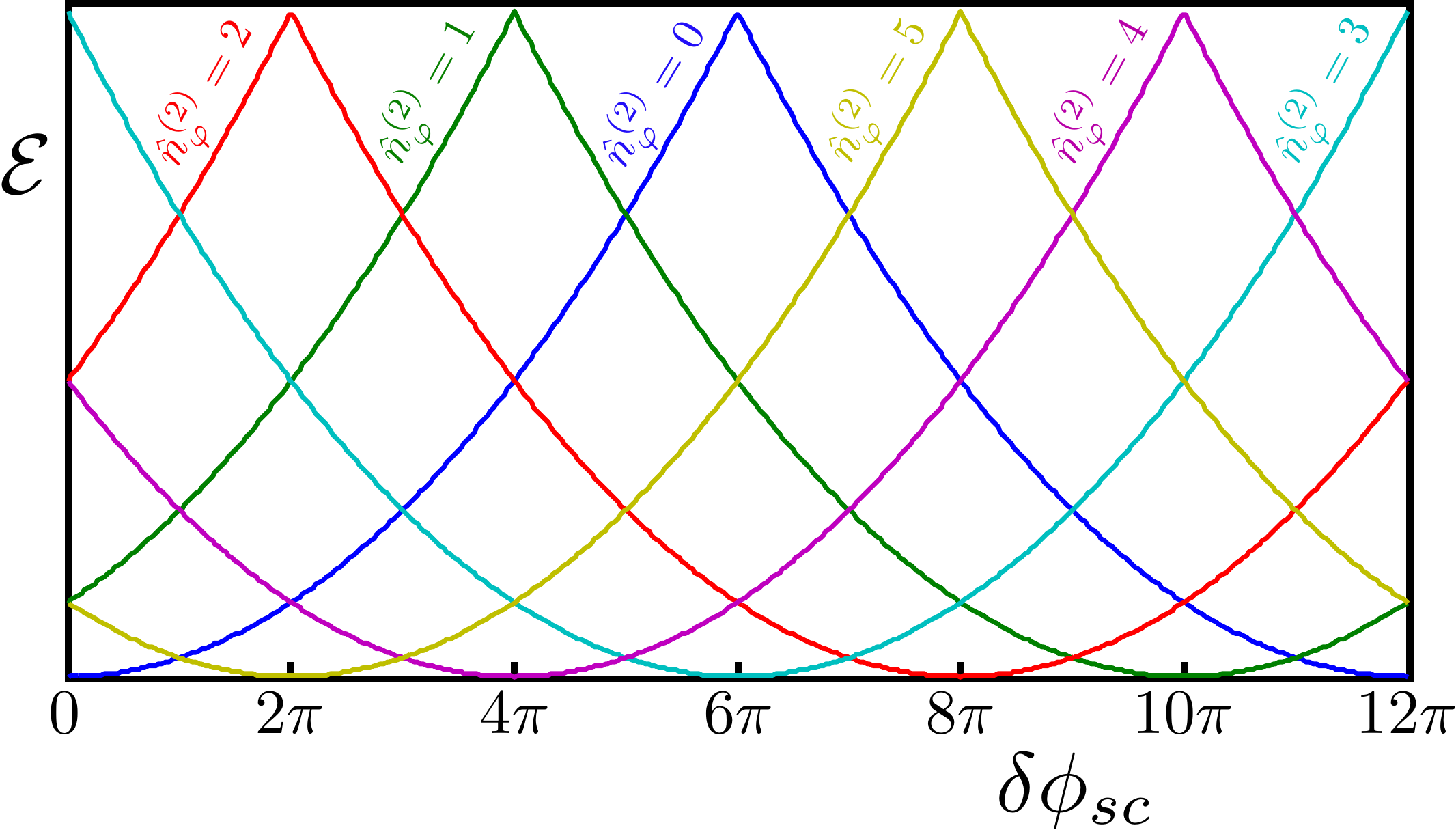}
\caption{Energy versus superconducting phase difference $\delta\phi_{sc}$ across the Josephson junction in the $m = 3$ case.  The six curves shown correspond to the distinct values of $\hat{n}_\varphi^{(2)}$ characterizing the pinning of $\varphi$ under the right superconductor, assuming that $\varphi = 0$ beneath the left superconductor.  Provided $\hat{n}_\varphi^{(2)}$ is conserved the energy and hence the current are both $12\pi$ periodic in $\delta\phi_{sc}$.}
\label{energy_vs_phase}
\end{figure}

Since $[H,\hat{n}^{(2)}_\varphi] = 0$ the eigenvalue of $\hat{n}^{(2)}_\varphi$ is a conserved quantity that can not change under adiabatic evolution of the Hamiltonian.  It is this crucial property that gives rise to `fractional' Josephson effects.  \emph{For a fixed initial value of} $\hat{n}^{(2)}_\varphi$, one sees from Eqs.\ (\ref{varphiBC}) and (\ref{E}) that the energy is $4\pi m$ periodic in $\delta\phi_{sc}$, despite the fact that the underlying Hamiltonian---recall Eq.\ (\ref{H1p})---clearly exhibits $2\pi$ periodicity.  [Note that here is where compactness of $\varphi$ is essential.  Had we expressed $\varphi(x_2 + \ell)$ in Eq.\ (\ref{varphiBC}) without modding by $2\pi$ the energy would increase unboundedly with $\delta\phi_{sc}$, which is obviously physically incorrect.]  As a concrete illustration, Fig.\ \ref{energy_vs_phase} displays the energy $\mathcal{E}(\delta\phi_{sc})$ versus $\delta\phi_{sc}$ for the six inequivalent $\hat{n}^{(2)}_\varphi$ values in the $m = 3$ case.

As mentioned in the main text the Josephson current flowing across the junction exhibits the same $4\pi m$ periodicity as the energy.  One should, however, bear in mind the following caveats that have been raised in the context of the Majorana-mediated fractional Josephson effect (see, \emph{e.g.}, Refs.\ \onlinecite{Kitaev2001,Fu2009a}).  In any experiment the measured current will consist of a $4\pi m$-periodic contribution arising from the fused parafermions \emph{and} a conventional $2\pi$-periodic component flowing in parallel.  (The latter can arise, for example, from the ordinary Josephson current that flows directly between the two parent $s$-wave superconductors.)  These currents must be disentangled if one is to utilize Josephson measurements to read out the qubits encoded by the parafermions.  We also note that in practice various imperfections---\emph{e.g.}, inelastic processes that change the value of $\hat{n}^{(2)}_\varphi$ or additional parafermion couplings that spoil conservation of $\hat{n}^{(2)}_\varphi$---can potentially restore $2\pi$ periodicity of the current.  Exploring these subtleties in detail would be quite interesting, particularly given the fractionalized nature of the system we are dealing with.

\section{Quasiparticle tunneling at the junction in Fig.\ \ref{FQH_fig2}(b)}
\label{TunnelingAppendix}

Section \ref{sec-trans} noted that at the junction in Fig.\ \ref{FQH_fig2}(b) charge $e/m$ excitations can tunnel between the right-moving (red) edge states, whereas only electrons can tunnel between the left-moving (blue) edge states.  The distinction between these allowed processes is crucial for the outcome of the braid analyzed there.  We now wish to elaborate on this point by examining the junction Hamiltonian in greater detail.  Let $-x_0$ and $+x_0$ respectively denote the coordinates of the left and right sides of the junction where tunneling takes place.  One can then model the coupling of right- and left-moving modes across the sack by
\begin{eqnarray}
  H_{J} &=& -t_R\cos[\phi_R(x_0)-\phi_R(-x_0)+\beta_R]
  \nonumber \\
  &-& t_L\cos[m(\phi_L(x_0)-\phi_L(-x_0))+\beta_L].
\end{eqnarray}
Here $t_R$ is the tunneling amplitude for $e/m$ right-moving excitations, $t_L$ is the tunneling amplitude for left-moving electrons, and $\beta_{R/L}$ are non-universal phases.  Higher-order processes such as pair tunneling can be easily incorporated in what follows, but we neglect these for simplicity.

Our primary interest here is to understand how $H_{J}$ couples parafermion modes.  For concreteness let us consider the setup in Fig.\ \ref{FQH_fig2}(d) where zero-modes $\alpha_2$ and $\alpha_1'$ reside on opposite sides of the junction.  Due to the pinning of $\varphi$ and $\theta$ near the domain walls in the figure, one can replace
\begin{eqnarray}
  \phi_{R/L}(-x_0) &=& \frac{\pi}{m}(\hat{n}_\varphi^{(1)}\pm\hat{n}_\theta^{(2)})
  \\
  \phi_{R/L}(x_0) &=& \frac{\pi}{m}(\hat{n}_\varphi^{(2)}\pm\hat{n}_\theta^{(3)}),
\end{eqnarray}
yielding
\begin{eqnarray}
  H_{J} &=& -t_R\cos\left[\frac{\pi}{m}(\hat{n}_\varphi^{(2)}+\hat{n}_\theta^{(3)}-\hat{n}_\varphi^{(1)} - \hat{n}_\theta^{(2)}) + \beta_R\right]
  \nonumber \\
  &-& t_L\cos[\pi(- \hat{n}_\varphi^{(2)}+\hat{n}_\theta^{(3)} + \hat{n}_\varphi^{(1)} - \hat{n}_\theta^{(2)})+\beta_L].
\end{eqnarray}
The first line simply corresponds to the $(t_J\alpha_2^\dagger \alpha_1' + H.c.)$ parafermion hybridization term in Eq.\ (\ref{Hcd}).  Importantly, the second line is a function of the same linear combination of operators $\hat{n}_\varphi^{(2)}+\hat{n}_\theta^{(3)}-\hat{n}_\varphi^{(1)} - \hat{n}_\theta^{(2)}$ that appears in $t_R$ above.  Consequently electron tunneling results in a benign coupling of the form $[t_J'(\alpha_2^\dagger\alpha_1')^m + H.c.]$ that does not change the conclusions in Sec.\ \ref{sec-trans} (apart, perhaps, from an unimportant modification of the non-universal parameter $k$ appearing in the braid transformation).  The factor of $\pi$ in the cosine---which reflects the fact that $t_L$ describes tunneling of electrons rather than charge $e/m$ quasiparticles---underlies this important result.  If fractionalized quasiparticles could tunnel between right- \emph{and} left-movers, then additional non-local terms would appear involving parafermions far from the junction.  For example, tunneling of left-moving $e/m$ quasiparticles would produce a term of the form $[\delta t (\alpha_2'^2)^\dagger\alpha_1'\alpha_2 + H.c.]$.  Such couplings, when combined with right-moving $e/m$ tunneling, would spoil the topological nature of the braids we analyzed but fortunately are precluded in our system.

\section{Controlled-phase gate from sequential braids}\label{braidapp}

In this final appendix we consider sequential braids that produce a controlled-phase gate.  For convenience, we relabel the ground state basis in terms of
\begin{equation}
 \ket{q}'_k= \ket{q+k+m}
\end{equation}
so that Eq.\ (\ref{U12}) yields simply
\begin{equation}
U_{12}\ket{q}'_k=e^{-i\frac{\pi}{2m} q^2}\ket{q}'_k.
\end{equation}
Henceforth we drop the label $k$, assuming that all exchanges occur through a single junction characterized by the same $k$.

Using this definition, we now examine the effects of more complicated braids.  As a concrete example consider Fig.\ \ref{FQH_fig2}(b) where two pairs of domain walls bind four parafermion zero-modes $\alpha_{1,\ldots,4}$, yielding a ground state space with $(2m)^2$ states (assuming no overall fusion channel constraint).  For ease of notation let us refer to the domain walls binding $\alpha_j$ simply by $j$.  One can implement a clockwise exchange of the left and right pairs of domain walls via the individual clockwise exchanges of 2 and 3, followed by 3 and 4, 1 and 2, then 2 and 3 once more, building up the unitary $U=U_{23}U_{12}U_{34}U_{23}$.
In order to determine the effects of this braid, we first fix the relative phases of the ground states by defining $\ket{p,q}'=\alpha_1^{\dagger p}\alpha_3^{\dagger q}\ket{0}'\otimes\ket{0}'$, where the first term in the direct product denotes the combined state of parafermions $\alpha_1$ and $\alpha_2$, while the second denotes the combined state of parafermions $\alpha_3$ and $\alpha_4$.  Note that with our conventions we have
\begin{equation}
\alpha_1^\dagger\alpha_2\ket{0}'=-e^{i\frac{\pi}{m}(k+m-1/2)}\ket{0}',
\end{equation}
and similarly for $\alpha_3^\dagger\alpha_4$.

The total effect of $U$ is to transform
\begin{eqnarray}
\alpha_1&\rightarrow& e^{2\pi i (k-1)/m}\alpha_3\nonumber\\
\alpha_2&\rightarrow& e^{2\pi i (k-1)/m}\alpha_4\nonumber\\
\alpha_3&\rightarrow& \alpha_3^{\dagger 2}\alpha_1\alpha_4^2\nonumber\\
\alpha_4&\rightarrow& \alpha_3^{\dagger 2}\alpha_2\alpha_4^2.
\end{eqnarray}
In particular, $U\alpha_1^\dagger\alpha_2U^\dagger=\alpha_3^\dagger\alpha_4$ and $U\alpha_3^\dagger\alpha_4U^\dagger=\alpha_1^\dagger\alpha_2$, which implies that $U\ket{p,q}'=e^{i\kappa_{pq}}\ket{q,p}'$.  The above relations allow one to determine the phases $\kappa_{pq}$; upon discarding an overall phase we obtain
\begin{equation}
U\ket{p,q}'=e^{i\frac{\pi}{m}\left[2k(p-q)-pq\right]}\ket{q,p}'.
\end{equation}
Double exchange of these two pairs of domain walls thus yields, again up to an overall phase,
\begin{equation} 
  U^2\ket{p,q}'=e^{-i\frac{2\pi}{m} pq}\ket{p,q}'.
  \label{U2}
\end{equation}
corresponding to the $CP$ gate from Eq.\ (\ref{CP}) upon transforming back to our original basis.  (Note that the junction parameter $k$ cancels out here.)  
Equation (\ref{U2}) constitutes a rather important result: because the phase factor on the right side depends on both $p$ and $q$, this braid operation can entangle the two registers when acting on a superposition of orthogonal ground states.  This braid distinguishes parafermions from Majoranas, since $U^2$ is trivial when $m=1$.

The above braids, together with the $(2m)^\mathcal{N}$ fold degeneracy of the ground state manifold, suggest the following set of fusion rules for the parafermion modes:
\begin{eqnarray}
\alpha\otimes\alpha&=& \psi_0 \oplus \psi_1 \oplus \cdots \oplus \psi_{2m-1}
\\
\alpha\otimes\psi_j&=&\alpha
\\
\psi_{j_1}\otimes\psi_{j_2}&=&\psi_{j_1+j_2~ \text{mod}~ 2m}.
\end{eqnarray}
Here $\psi_0$ is the identity channel and $\psi_{1,\ldots,2m-1}$ represent the distinct quasiparticle types to which pairs of parafermions can fuse.  (The first line can be intuitively understood from our analysis of the parafermionic Josephson effect.)  The parafermion $\alpha$ has quantum dimension $\sqrt{2m}$, while each other field has dimension 1. Note that the set of $\psi_j$ form an Abelian sub-algebra consistent with the braid operator $U$ found above. For $m=1$, these fusion rules reduce to the well-known Ising anyon theory.


\end{document}